\documentclass[aps,nofootinbib,showpacs,showkeys,preprint
,tightenlines ] {revtex4}

\usepackage{epsf,epsfig,subfigure,graphicx,amsmath,amssymb}
\usepackage{color}
\usepackage{latexsym}
\usepackage{hyperref}
\usepackage{bm}
\usepackage{ifthen}

\newcommand{\dis}[1]{\begin{equation}\begin{split}#1\end{split}\end{equation}}
\newcommand{\be}{\begin{equation}}
\newcommand{\ee}{\end{equation}}

\newcommand{\eq}[1]{Eq.~(\ref{#1})}

\newcommand{\bfrac}[2]{{\left(\frac{#1}{#2} \right)  }}\newcommand{\VEV}[1]{\langle #1 \rangle}
\newcommand{\Mp}{M_P}

\newcommand{\baru}{{\overline{u}}}
\newcommand{\barv}{{\overline{v}}}

\newcommand{\p}{{\partial}}
\newcommand{\vp}{\varphi}
\newcommand{\vps}{\varphi_*}
\newcommand{\vpe}{\varphi_e}
\newcommand{\chis}{\chi_*}
\newcommand{\chie}{\chi_e}

\newcommand{\fnl}{f_{\rm NL}}

\newcommand{\Ps}{\mathcal{P}}

\def\bkone{{\bf k_1}}
\def\bktwo{{\bf k_2}}

\def\picube{(2\pi)^3}
\newcommand{\sdelta}[1]{\!\delta^{\,3}(\mathbf{#1})}

\newcommand{\sgn}{{\rm sign}}

\newcommand{\bcalAp}{\overline{{\cal A}_P}}
\newcommand{\bcalAs}{\overline{{\cal A}_S}}

\newcommand{\calP}{{\cal P}}

\def\etap{\eta_{\varphi\varphi}}
\def\etac{\eta_{\chi\chi}}
\def\etape{\eta^e_{\varphi\varphi}}
\def\etace{\eta^e_{\chi\chi}}
\def\etaps{\eta^*_{\varphi\varphi}}
\def\etacs{\eta^*_{\chi\chi}}
\def\epspe{\epsilon_{\varphi}^e}
\def\epsps{\epsilon_{\varphi}^*}
\def\epsce{\epsilon_{\chi}^e}
\def\epscs{\epsilon_{\chi}^*}
\def\bea{\begin{eqnarray}}
\def\eea{\end{eqnarray}}

\def\beps{\overline{\epsilon}}
\def\bepse{\overline{\epsilon^e}}
\def\bu{\overline{u}}
\def\bv{\overline{v}}

\begin{document}
%\draft

\begin{flushright}
{\tt  APCTP-Pre2012-002
\\PNUTP-12-A02}
\end{flushright}

\title{\bf \Large Primordial curvature perturbation 
\\during and at the end of multi-field inflation}

\author{ Ki-Young Choi$^{(a)}$$^{(b)}$\footnote{email: kiyoung.choi@apctp.org}, Soo A Kim$^{(a)}$\footnote{email: sooastar@gmail.com},
and Bumseok Kyae$^{(c)}$\footnote{email: bkyae@pusan.ac.kr} }
\affiliation{$^{(a)}$ Asia Pacific Center for Theoretical Physics, Pohang, Gyeongbuk 790-784, Republic of Korea\\
$^{(b)}$Department of Physics, POSTECH, Pohang, Gyeongbuk 790-784, Republic of Korea\\
$^{(c)}$ Department of Physics, Pusan National University, Busan
609-735, Republic of Korea}

%\maketitles

\begin{abstract}
We study the generation of the primordial curvature perturbation in multi-field inflation.  Considering both the evolution of the perturbation during inflation and the effects generated at the end of inflation, we present a general formula for the curvature perturbation. 
We provide the analytic expressions of the power spectrum, spectral tilt and non-Gaussianity
for the separable potentials of two inflaton scalars, and apply them to some specific models.
\end{abstract}

%\pacs{\cred{98.80.Cq, 12.60.Jv, 04.65.+e}}

 \maketitle

%%%%%%%%%%%%%%%%%%%%%%%%%%%%%%%%%%%%%%%%%%%%%%%%%%%%%%%%%%%%%%%%%%%%%%%%%%%%%%%%%
%%%%%%%%%%%%%%%%%%%%%%%%%%%%%%%%%%%%%%%%%%%%%%%%%%%%%%%%%%%%%%%%%%%%%%%%%%%%%%%%%

\section{introduction}

The generation of the large scale structures and the anisotropy in the temperature of the cosmic microwave background  (CMB) suggests that there was already small inhomogeneity in the early Universe, a few Hubble times before the observable scale enters the horizon.  The time-independent  curvature perturbation $\zeta$ sets the initial conditions for such inhomogeneity and the subsequent evolution of  all the scalar perturbations.  The resulting power spectrum and the  spectral tilt of the primordial curvature perturbation are almost Gaussian, and so scale-independent with the size of ${\cal P}_\zeta = 2.43\times 10^{-9}$~\cite{Komatsu:2010fb}. 

The perturbations necessary for the inhomogeneity  can arise naturally from the vacuum fluctuations 
of light scalar field(s) during inflation and be promoted to classical one around the time of the horizon exit. In the single field inflation with the canonical kinetic term,  the curvature perturbations produced at the horizon crossing are conserved after that and can explain the primordial curvature perturbation necessary for the observation.

Inflation with multiple scalar fields~\cite{Bassett:2005xm,Wands:2007bd}, however, can admit quite different inflationary dynamics and spectra of the primordial perturbations, which is  impossible in single field inflationary models. The presence of multiple field induces non-adiabatic field perturbations and  the evolution of the overall curvature perturbations during inflation, possibly leading to detectable non-Gaussianity~\cite{Byrnes:2008wi}.
The evolution of the curvature perturbation continues until the non-adiabatic perturbation is converted to the adiabatic one. The conversion must be completed before the cosmologically relevant scales enter the horizon again  in the deep radiation-dominated era after inflation, and thus we need to track the evolution of the curvature perturbation until it becomes frozen. 

The other dominant component of the curvature perturbation in the multi-field inflation is generated at the transition between inflation and non-inflation phase~\cite{Lyth:2005qk}, which cannot happen in the single field inflation. In the single component inflation, the inflationary trajectory  is unique and the inflation
ends when the inflaton field $\vp$ has a value $\vp_e$, which is controlled entirely by the inflation and independent of the position. With additional scalar fields, however, the trajectory is in the multi-dimensional field space and the inflation ends with different field values.  Then  the field values of the inflatons depend on the positions and the field space at the end of inflation is not in the uniform energy density any more.  The relative differences of ending inflation make additional differences in the e-folding number and contribute to the curvature perturbations at the end of inflation.

A simple example of generating the curvature perturbation at the end of inflation is the hybrid inflation with sudden-end approximation~\cite{Lyth:2005qk,Lyth:2006nx}. In this case, 
the end of inflation occurs suddenly by the dynamics of waterfall fields, and  the observable large non-Gaussianity can be generated. The realization in the string theory was considered in~\cite{Lyth:2006nx}.
A specific analytic calculation was done for two-component hybrid inflations in Ref.~\cite{ Sasaki:2008uc,Naruko:2008sq,Huang:2009xa} with an exponential potential, and they also found that large non-Gaussianity is possible for certain conditions at the end of inflation.
The more general calculation was attempted in Ref.~\cite{Byrnes:2008zy} for two-field hybrid inflation. They studied the generation of  large non-Gaussianity both during the inflation and at the end of inflation. They found that the end condition of inflation can change the pre-existing large non-Gaussianity severely by changing the sign or the magnitude.

The generation of curvature perturbation can be understood geometrically. In ~\cite{Huang:2009vk},  for simplicity, the straight trajectory was considered and the relation between the non-Gaussianity and the geometrical properties for the hyper-surface of the end of inflation  were studied.
The generation of the primordial statistical anisotropy at the end of inflation was also considered in~\cite{Yokoyama:2008xw} and~\cite{Emami:2011yi}.

Both the evolution of the curvature perturbation during inflation and the contribution at the end of inflation are the general properties in multi-field inflation models. To get the correct 
curvature perturbation after inflation, it is necessary to consider both of them in a consistent way.
In this paper, we propose the general formula for the curvature perturbation after the end of inflation,
considering both effects 
 in the case of the separable potentials by product or sum. Here we will use  $\delta N$ formalism and provide analytic formulae for the power spectrum, spectral index and non-Gaussianity parameters.

This paper is organized as follows. In Section~\ref{basic}, we present the basic formulae, and in Section~\ref{perturbation} we describe the background for our calculation and provide the analytic formulae for curvature perturbation. They are our main results. In Section~\ref{application}, we apply the formulae derived in the previous section to calculate the power spectrum and non-linearity parameters
in the several models
and conclude in Section~\ref{conclusion}.

%%%%%%%%%%%%%%%%%%%%%%%%%%%%%%
%										%
%				Background theory		         %
%										%
%%%%%%%%%%%%%%%%%%%%%%%%%%%%%%
\section{Background theory}
\label{basic}

We consider an inflationary scenario implemented by multi-inflaton scalars $\phi_1,\phi_2,\cdots,\phi_n$ ($\equiv\phi_I$) with the {\it canonical} kinetic terms and the potential $W(\phi_1, \phi_2, \cdots,\phi_n)$. The classical background of the Universe are governed by the Friedmann equation and the equations of motion for the inflaton fields. In the slow-roll regime, the Friedmann and field equations are approximately given by 
\dis{
3\Mp^2 H^2 \simeq W\, ,~~
3H\dot\phi_I \simeq -\frac{\partial W}{\partial \phi_I}\,.
}
The slow-roll parameters are defined as 
\begin{eqnarray}
&&\epsilon_{\phi_I}
 \equiv \frac{\Mp^2}{2}\left[\frac{1}{W}\left( \frac{\partial W}{\partial \phi_I}\right)\right]^2
  \equiv \frac{\Mp^2}{2} \bfrac{W_{\phi_I}}{W}^2\,,\\
&&\eta_{\phi_I \phi_J} 
  \equiv \Mp^2 \left[\frac{1}{W}\frac{\partial^2 W}{\partial \phi_I \p \phi_J}\right]
  \equiv  \Mp^2 \frac{W_{\phi_I \phi_J}}{W}\,. \label{slow-roll}
\end{eqnarray}

The curvature perturbation $\zeta$ on super horizon scales can be evaluated using $\delta N$-formalism~\cite{starob85,ss1,Sasaki:1998ug,lms}, which is the perturbation of the e-folding number $N$ defined as
\dis{
N(t_c,t_*,{\bf x}) \equiv \int_*^c H dt . \label{efoldingN}
}
It is evaluated from an initial flat hypersurface at $t=t_*$ to a final uniform density hypersurface
at $t=t_c$.  Here we take $t_*$ as the time of the horizon exit of the relevant scale during inflation
and $t_c$ as some time later after inflation.
Then the number of e-foldings $N$ can be a function of the field configuration $\phi_I(t_*,{\bf x})$ on the flat hypersurface at $t=t_*$ and the time of $t_c$.
Then the perturbation of the number of e-foldings $\delta N (t_c,t_*,{\bf x})$ can be expanded in terms of the field perturbations $\delta \phi_I(t_*,{\bf x})$.
The curvature perturbation  is given by
\dis{\label{deltaN} 
\zeta\simeq \delta N= \sum_I
N_{,I}\delta\phi_{I*}+\frac12\sum_{IJ}N_{,IJ}\delta\phi_{I*}\delta\phi_{J*}
}
up to quadratic terms, where the first and second derivatives are 
\dis{
N_{,I}\equiv \frac{\p N}{\p \phi_*^I},\qquad N_{,IJ}\equiv \frac{\p^2 N}{\p \phi_*^I \p\phi_*^J} .
\label{efoldingNij}
}
Here we assumed the slow-roll conditions and ignored the possible dependence on their first time derivatives $\dot{\phi}_I(t)$.
The field perturbations $\delta \phi_{I*}$ satisfy the two-point correlation function,
\dis{
\VEV{\delta \phi_{I*}({\bf k_1}) \delta \phi_{J*}({\bf k_2})} = (2\pi)^3 \delta_{IJ}^{(3)} ( {\bf k_1}  + {\bf k_2} ) \frac{2\pi^3}{k_1^3} {\cal P}_*(k_1),\qquad  {\cal P}_*(k)\equiv\frac{H_*^2}{4\pi^2},
}
where $H_*$ is evaluated at Hubble exit, $k=aH$.

The power spectrum of the curvature perturbation $\zeta$, $\calP_{\zeta}$, and the bispectrum $B_\zeta$ are defined as
\begin{eqnarray}\label{powerspectrumdefn} \langle\zeta_{\bkone}\zeta_{\bktwo}\rangle &\equiv&
\picube\,
\sdelta{\bkone+\bktwo}\frac{2\pi^2}{k_1^3}\calP_{\zeta}(k_1) \, , \\
\langle\zeta_{{\mathbf k_1}}\,\zeta_{{\mathbf k_2}}\,
\zeta_{{\mathbf k_3}}\rangle &\equiv& \picube\, \sdelta{{\mathbf
k_1}+{\mathbf k_2}+{\mathbf k_3}} B_\zeta( k_1,k_2,k_3) \,. \end{eqnarray}
Observational limits are usually put on the non-linearity parameter $\fnl$, which is defined by~\cite{Maldacena:2002vr}
\dis{
\fnl=\frac56\frac{k_1^3k_2^3k_3^3}{k_1^3+k_2^3+k_3^3}
\frac{B_{\zeta}(k_1,k_2,k_3)}{4\pi^4\calP_{\zeta}^2}. \label{fnldefn}
}

Using \eq{deltaN}, one obtains the power spectrum of the curvature perturbation as
\dis{
\calP_\zeta = \sum_I N_{,I}^2 \calP_*. \label{Pzeta}
}
From this we can define the observable quantities, the spectral index and the
tensor-to-scalar ratio:
\bea &&n_{\zeta}-1\equiv \frac{\partial \log\calP_{\zeta}}{\partial\log k}, \label{tilt}\\
&&r=\frac{\calP_T}{\calP_{\zeta}}=\frac{8\calP_*}{\Mp^2\calP_{\zeta}}, 
 \eea
where $\calP_T=8\calP_*/\Mp^2=8H_*^2/(4\pi^2\Mp^2)$ is the power spectrum of
the tensor metric fluctuations.
The non-linearity parameter $\fnl$ is given by~\cite{Lyth:2005fi}
\dis{
\fnl= \frac56 \frac{\sum_{IJ}N_{,I}N_{,J}N_{,IJ}}{\left(\sum_K N_{,K}N_{,K}\right)^2}.
\label{fnl}
}

In \eq{efoldingN}, the final time can be identified as the end of inflation, if the field values at the end of inflation are on the uniform energy density hypersurface. In multi-field inflation, however, that does not happen in general; the inflation ends at slightly different times in different places.
This inhomogeneous end of inflation can generate curvature perturbation additionally~\cite{Lyth:2005qk}.
To include the effect from the end of inflation, we extend the final time $t_c$ to  sometime in the deep radiation dominated era after inflation, 
while we denote $t_e$ as the time at the end of inflation. Thus, the e-folding number has the two contributions; one is from inflation phase $N_e(t_*,t_e)$ and the other from  radiation phase $N_c(t_e,t_c)$, 
\dis{
N(t_*,t_c) = N_e(t_*,t_e) + N_c(t_e,t_c).
}
Here, we take into account a sudden change from the inflation phase to the radiation era.
The change in the e-folding number during the radiation dominated era between $t_e$ and $t_c$ is given by ~\cite{Byrnes:2008zy}
\dis{
N_c=\frac14 \log \bfrac{W_e}{W_c}\label{Nc},
}
where $W_e$ is the potential energy as a function of the field values at the end of inflation and $W_c$ is 
the energy density at a time $t_c$ during the radiation dominated era,  
which is uniform in space.

Thus, $\delta N$ has the two contributions from the inflation and radiation epochs,
\dis{
\delta N = \delta N_e + \delta N_c, \label{deltaNec}
}
with 
\dis{
\delta N_c = \frac{1}{4W_e}\sum_I \frac{\p W_e}{\p \phi_I^*} \delta \phi_I^* \label{deltaNc}.
}
Here $\delta N_c$ depends only on the epoch at the transition between inflation and radiation;
it does not depend on the final time.
The reason is manifest since the curvature perturbation  is conserved in the radiation dominated era~\cite{lms}.

Using the slow-roll parameters, \eq{slow-roll},  $\delta N_c$ becomes 
\dis{
\delta N_c =\sum_{IJ} \frac{\sqrt{2\epsilon^e_{\phi_J}}}{4\Mp} \bfrac{\p \phi_J^e}{\p \phi_I^*} \delta \phi_I^*,
\label{deltaNcSR}
}
where the  superscript ${}^e$ denote the values evaluated at the end of inflation $t=t_e$.
As we will see later, $\delta N_c$ is suppressed compared to $\delta N_e$ by the slow-roll parameters.

%%%%%%%%%%%%%%%%%%%%%%%%%%%%%%
%										%
%				Curvature perutrbations  		%
%										%
%%%%%%%%%%%%%%%%%%%%%%%%%%%%%%
%%%%%%%%%%%%%%%%%%%%%%%%%%%%%%%%%%%%%%%%%%%%%%
\section{Curvature perturbation generated during and at  the end of inflation}
\label{perturbation}
%%%%%%%%%%%%%%%%%%%%%%%%%%%%%%%%%%%%%%%%%%%%%%

In this section, we will discuss the contribution to curvature perturbation from inflationary era $\delta N_e$ and $\delta N_c$ in \eq{deltaNec} with general condition of ending inflation. We attempt to get analytic formulae for the power spectrum ($\cal{P}_\zeta$), the spectral index ($n_\zeta$), and the non-linearity parameter ($f_{NL}$) particularly in the cases that the inflaton potential is given by the product, $W(\vp,\chi)=U(\vp)V(\chi)$, or by sum,  $W(\vp,\chi)=U(\vp)+V(\chi)$. Since these forms of the potentials cover the large class of inflation models, the analytic expressions for $\cal{P}_\zeta$, $n_\zeta$, and $f_{NL}$ would be very useful.

%%%%%%%%%%%%%%% 
% 		 Product		    %
%%%%%%%%%%%%%%%
\subsection{Product potential $W(\vp,\chi)=U(\vp)V(\chi)$} 
In this subsection,  we consider the case of a separable potential by product, 
\dis{
W(\vp,\chi)=U(\vp)V(\chi)  \label{potentialproduct}.
}
First, we derive the curvature perturbation, and then from this we obtain the analytic formulae for the power spectrum $\cal{P}_\zeta$ and the non-linear parameter $\fnl$. 
We assume the slow-roll during inflation. From \eq{slow-roll}, 
the slow-roll parameters are  given by\begin{eqnarray}
&&\epsilon_\vp
  = \frac{\Mp^2}{2}\left(\frac{U_\vp}{U}\right)^2, \qquad
\epsilon_\chi
  = \frac{\Mp^2}{2}\left(\frac{V_\chi}{V}\right)^2,\\
&&~\etap
  =\Mp^2 \frac{U_{\vp\vp}}{U} ~,~\qquad\quad
 \etac
  =\Mp^2 \frac{V_{\chi\chi}}{V} ~,
\end{eqnarray}
and supposed to be small.

The number of e-foldings with the potential in \eq{potentialproduct} in the slow-roll regime is given by~\cite{GarciaBellido:1995qq}
\dis{
N_e(\vp_*,\chi_*) = -\frac{1}{\Mp^2} \int^e_*\frac{U}{U_\vp}d \vp = -\frac{1}{\Mp^2} \int^e_*\frac{V}{V_\chi}d \chi.  \label{Nproduct}
}
The integrals in \eq{Nproduct} depend on both field values at horizon crossing ($\vp_*$ and $\chi_*$) and at the end of inflation ($\vp_e$ and $\chi_e$). Note that  $\vp_e$ and $\chi_e$ also depend on the initial field values,  $\vp_*$ and $\chi_*$ along the given trajectory.  The differentiation of the number of e-foldings $dN_e$ is given by
\dis{
dN _e
=\frac{1}{\Mp^2} \left[  \bfrac{U}{U_\vp}_* -\bfrac{U}{U_\vp}_e \frac{\p \vp_e}{\p \vp_*}\right] d\vp_*  
- \frac{1}{\Mp^2} \bfrac{U}{U_\vp}_e\frac{\p \vp_e}{\p \chi_*}d\chi_*  . \label{dNproduct}
} 
The derivatives of the final field values with respect to the initial field values also depend on the condition 
of the fields at the end of inflation.
Here we suppose that the inflation ends when the fields satisfy the condition given by
\dis{
E(\vpe,\chie)= \text{a constant} \label{endcondition}.
}
This is different from that used in the previous analyses in Ref.~\cite{GarciaBellido:1995qq,Choi:2007su}, where the uniform energy density hypersurface is simply employed at the end of inflation.

With the condition \eq{endcondition}  at the end of inflation, we find that the derivatives are given by 
\dis{
\frac{\p \vpe}{\p\vps}=\frac{\epsce}{\bepse}\sqrt{\frac{\epspe}{\epsps}} ~, \qquad &
\frac{\p \vpe}{\p\chis}=-\frac{\epsce}{\bepse}\sqrt{\frac{\epspe}{\epscs}} ~,\\
\frac{\p \chie}{\p\vps}=-\frac{R\epspe}{\bepse}\sqrt{\frac{\epsce}{\epsps}} ~, \qquad &
\frac{\p \chie}{\p\chis}=\frac{R\epspe}{\bepse}\sqrt{\frac{\epsce}{\epscs}} ~, \label{dervsproduct}
}
where the superscript (or subscript) ${}^e$ and ${}^*$ denote the values evaluated at the end of inflation and horizon crossing, respectively. 
Throughout this paper it should be understood that there is additional factor of $\sgn(U_{\vp})$  or $\sgn(V_{\chi})$  whenever  square root  appears. We note that there is an additional factor $R$, which contains the information about the condition of ending inflation~\eq{endcondition}.
In \eq{dervsproduct}, $R$ and $\beps$ are defined by
\dis{
R &\equiv \left(\frac{U/U_\vp}{V/V_\chi}\right)_e\, \bfrac{\p E/\p \vpe}{\p E/ \p\chie}=\sqrt{\frac{\epsilon_\chi^e}{\epsilon_\vp^e}} \frac{E_\vp}{E_\chi},\qquad  \\
\bepse & \equiv R\epspe + \epsce=\frac{\sqrt{\epsilon_\chi}}{E_\chi}( E_\vp\sqrt{\epsilon_\vp} +E_\chi\sqrt{\epsilon_\chi} ), \label{R}
}
with $E_{\vp}\equiv\p E/ \p \vp_e$ and $E_{\chi}\equiv\p E/ \p \chi_e$.
Hence, if the condition for the end of inflation is given by the uniform energy condition, $E(\vp_e,\chi_e)=W(\vp_e,\chi_e)=$ a constant, then $E_\vp$ and $E_\chi$ are given by $VU_\vp|_e$ and $UV_\chi|_e$, respectively, which makes $R$ the unity.
Note that $R$ can be a {\it negative as well as positive} value, and so $\overline{\epsilon^e}$ has possibly a vanishing limit.

Using \eq{dNproduct} and ~\eq{dervsproduct},  we find the first derivatives of e-folding number $N(t_e,t_*)$ with respect to the fields,
\dis{
&\Mp\frac{\p N_e}{\p \vps}=\frac{\bu}{\sqrt{2\epsps}},\\
&\Mp\frac{\p N_e}{\p \chis}=\frac{\bv}{\sqrt{2\epscs}},\label{linearproduct}
}
where we defined
\dis{ \label{uvbar}
\bu \equiv \frac{R\epspe}{\bepse} = \frac{E_\vp\sqrt{\epsilon_\vp^e}}{E_\vp\sqrt{\epsilon_\vp^e} +E_\chi\sqrt{\epsilon_\chi^e} }  ,\qquad \bv\equiv \frac{\epsce}{\bepse}=\frac{E_\chi\sqrt{\epsilon_\chi^e}}{E_\vp\sqrt{\epsilon_\vp^e} +E_\chi\sqrt{\epsilon_\chi^e} } ,
}
which are evaluated at the end of inflation.

Based on the first derivatives \eq{linearproduct}, we can find the second derivatives,
\dis{
\Mp^2\frac{\p^2 N_e}{\p \vps^2}&=\left( 1-\frac{\etaps}{2\epsps} \right)\bu
- \frac{1}{\epsps}\bcalAp,\\
\Mp^2\frac{\p^2 N_e}{\p \chis^2}&=\left( 1-\frac{\etacs}{2\epscs} \right)\bv
- \frac{1}{\epscs}\bcalAp,\\
\Mp^2 \frac{\p^2 N_e}{\p\vps\p\chis}
  &=
	\frac{1}{\sqrt{\epsps\epscs}}\bcalAp,\label{secondproduct}
}
where
\dis{
\bcalAp 
	&\equiv - \frac{R\epspe\epsce}{\bepse^2}
		     \left[  \overline{\eta_s^e}
			   - \frac{2(1+R)\epspe\epsce}{\bepse}
			  + \Mp \frac{\epspe\epsce}{R\bepse}
				\left\{ \frac{1}{\sqrt{2\epspe}} \frac{\p R}{\p \vpe} 
					-  \frac{R}{\sqrt{2\epsce}} \frac{\p R}{\p \chie}
   			          \right\}
		    \right] ,\\
\overline{\eta_s^e}&\equiv \frac{\epsce \etape + R\epspe \etace}{\bepse}. \label{barAp}
}

The other contribution to $\zeta$ coming from $\delta N_c$ can be obtained from \eq{deltaNcSR} using \eq{dervsproduct},
\dis{
\delta N_c = \frac{(1-R)}{2\Mp}\frac{\epsilon_\vp^e\epsilon_\chi^e}{\overline{\epsilon^e}} \left[ \frac{\delta \vp_*}{\sqrt{2\epsilon_\vp^*}} - \frac{\delta \chi_*}{\sqrt{2\epsilon_\chi^*}} \right].
}
We note that this contribution  disappears with $R=1$.  It means that the hypersurface of the end of inflation coincides with the uniform energy density hypersurface.

At the leading order the curvature perturbation is
\dis{
\zeta &= \delta N_e+\delta N_c\\
&= \frac{1}{ \Mp \overline{\epsilon^e}} 
	\left[     \left\{ R +\frac12(1-R)\epsilon_\chi^e  \right\} \frac{\epsilon_\vp^e}{\sqrt{2\epsilon_\vp^*}} \delta \vp_*
		+\left\{ 1 - \frac12(1-R)\epsilon_\vp^e\right\} \frac{\epsilon_\chi^e}{\sqrt{2\epsilon_\chi^*}}   \delta \chi_*
	\right].
} 
We find that the contribution from $\delta N_c$ is suppressed by another slow-roll parameters
in general except some extreme cases of very large or small $R$.
Ignoring   $\delta N_c$, the curvature perturbation is  determined solely by the contribution from the inflation epoch, $\zeta\simeq \delta N_e$.
In this case, the power spectrum \eq{Pzeta} is given by
\dis{ \label{powerPrd}
\calP_\zeta 
   \simeq \frac{W_*}{24\pi^2\Mp^4} \left(  \frac{\bu^2}{\epsilon_\vp^*} + \frac{\bv^2}{\epsilon_\chi^*} \right) \,,
   }
and the spectral index $n_\zeta$ and the tensor-to-scalr ratio $r$ are, respectively, given by
\dis{
n_\zeta & \simeq 1 -2 \epsilon^*
     -   \frac{4}{  \bu^2/\epsilon_\vp^* + \bv^2/\epsilon_\chi^*}
         \left[   \bu^2\left(1-\frac{\eta_{\vp\vp}^*}{2\epsilon_\vp^*}\right)
                  +\bv^2 \left(1-\frac{\eta_{\chi\chi}^*}{2\epsilon_\chi^*} \right) \right] \,,
\\
r
  &\simeq \frac{16}{ \bu^2/\epsilon_\vp^* + \bv^2/\epsilon_\chi^*},
}
where $\epsilon^* \equiv  \epsilon_{\phi}^*+\epsilon_{\chi}^*$.
The non-linearity parameter $\fnl$ is
\dis{ \label{fnlPrd}
\frac65\fnl 
\simeq \frac{2}{(\bu^2/\epsps +\bv^2/\epscs)^2}\left[ \frac{\bu^3}{\epsps}\left(1- \frac{\etaps}{2\epsps} \right)  +  \frac{\bv^3}{\epscs}\left( 1-\frac{\etacs}{2\epscs} \right) - \left( \frac{\bu}{\epsps}-\frac{\bv}{\epscs} \right)^2\bcalAp \right].
}
Note that the forms of $\calP_\zeta$ and $\fnl$ are the same as those in Ref.~\cite{Choi:2007su}, but the definitions of $\bu$, $\bv$ and $\bcalAp$ are different from those in Ref.~\cite{Choi:2007su}, which contain the informations
about the end of inflation.
We will find the new features from the special ending of inflation in the next section.

%%%%%%%%%%%%%%%%%%%%%%%%%%%%%%
\subsection{Separable potential by sum}
In this subsection we consider a separable potential by sum~\cite{VW};
\begin{equation}
W(\vp,\chi) = U(\vp) + V(\chi) \,.
\end{equation}
We keep the general condition for the end of inflaiton Eq.~(\ref{endcondition}).
The slow-roll parameters are given by
\dis{
&\epsilon_\vp 
    = \frac{\Mp^2}{2}\left( \frac{U_\vp}{W} \right)^2 \,,~~~
\epsilon_\chi
   = \frac{\Mp^2}{2}\left( \frac{V_\chi}{W} \right)^2 \,,
\\
& \eta_{\vp\vp}= \Mp^2 \frac{U_{\vp\vp}}{W} , \,\,~~~~~~~
 \eta_{\chi\chi}= \Mp^2 \frac{V_{\chi\chi}}{W}\,.
}

In the slow-roll limit the number of $e$-foldings is 
\begin{equation}
N_e
   = -\frac{1}{\Mp} \int^c_* \frac{U}{U_\vp}d\vp
	    -\frac{1}{\Mp} \int^c_* \frac{V}{V_\chi}d\chi \,,
\end{equation}
and its absolute derivative is given by
\dis{
dN_e
  = \frac{1}{\Mp}&
	\left[  \bfrac{U}{U_\vp}_* 
		- \frac{\partial\vp_e}{\partial \vp_*} \bfrac{U}{U_\vp}_e 
		- \frac{\partial \chi_e}{\partial \vp_*} \bfrac{V}{V_\chi}_e 
          \right] d\vp_* \\
        \qquad  & +
       \frac{1}{\Mp}
	\left[  \bfrac{V}{V_\chi}_* 
		- \frac{\partial\chi_e}{\partial \chi_*} \bfrac{V}{V_\chi}_e 
		- \frac{\partial \vp_e}{\partial \chi_*} \bfrac{U}{U_\vp}_e 
          \right] d\chi_* \,.
}
The field derivative terms at $t_*$ are 
\begin{eqnarray}
\frac{\partial \vp_e}{\partial \vp_*}
	= 	\frac{W_e}{W_*}\frac{\epsilon^e_\chi}{\bepse}\sqrt{\frac{\epsilon^e_\vp}{\epsilon^*_\vp}}\,,~~
\frac{\partial \vp_e}{\partial \chi_*}
	=     -\frac{W_e}{W_*}\frac{\epsilon^e_\chi}{\bepse} \sqrt{\frac{\epsilon^e_\vp}{\epsilon^*_\chi}}\,,~~\\
\frac{\partial \chi_e}{\partial \vp_*}
	=    -\frac{W_e}{W_*}\frac{R\epsilon^e_\vp}{\bepse}\sqrt{\frac{\epsilon^e_\chi}{\epsilon^*_\vp}}\,,~~
\frac{\partial \chi_e}{\partial \chi_*}
	= 	\frac{W_e}{W_*}\frac{R\epsilon^e_\vp}{\bepse}\sqrt{\frac{\epsilon^e_\chi}{\epsilon^*_\chi}}\,,
\end{eqnarray}
Similar to \eq{R}, $R$ and $\overline{\epsilon}$ are defined by
\dis{ \label{Rsum}
&R= \left(\frac{V_\chi}{U_\vp}\right)_e\left(\frac{E_\vp}{E_\chi}\right) ,\\
&\overline{\epsilon^e} = R \epsilon^e_\vp + \epsilon^e_\chi \,,
}
for the sum potential.
Hence, if the condition for the end of inflation is given by the uniform energy density condition, $E(\vp_e,\chi_e)=W(\vp_e,\chi_e)=$ a costant, then $E_\vp=U_\vp|_e$ and $E_\chi=V_\chi|_e$. So $R$ becomes again the unity.

Using these equaions, the first derivatives of the e-folding number with respect to the two inflaton fields at horizon exit are given by 
\dis{  \label{uv}
\Mp \frac{\partial N_e}{\partial \vp_*}
	=&	\frac{1}{\sqrt{2\epsilon^*_\vp}}\left( \frac{U_* + \widetilde{Z}_e}{W_*} \right)
	\equiv
		\frac{\bu}{\sqrt{2\epsilon^*_\vp}}\,,\\
\Mp \frac{\partial N_e}{\partial \chi_*}
	=&	\frac{1}{\sqrt{2\epsilon^*_\chi}} \left( \frac{V_* - \widetilde{Z}_e}{W_*} \right)
	\equiv  
			\frac{\bv}{\sqrt{2\epsilon^*_\vp}} \,,
}
where
\begin{equation} \label{Ztilde}
\widetilde{Z}_e 
	\equiv \frac{V_e R \epsilon^e_\vp - U_e \epsilon^e_\chi}{\bepse}\,.
\end{equation}
The second derivatives are given by
\begin{eqnarray}
\Mp^2 \frac{\partial^2 N_e}{\partial \vp_*^2}
	&=& 1-\frac{\eta_{\vp\vp}^*}{2\epsilon_\vp^*}~\bu
		+ 	\frac{\Mp}{W_*\sqrt{2\epsilon^*_\vp}}\frac{\partial\widetilde{Z}_e}{\partial \vp_*}\,,\\
\Mp^2 \frac{\partial^2 N_e}{\partial \chi_*^2}
	&=& 1-\frac{\eta_{\chi\chi}^*}{2\epsilon_\chi^*}~\bv
		- \frac{\Mp}{W_*\sqrt{2\epsilon^*_\chi}}\frac{\partial\widetilde{Z}_e}{\partial \chi_*}\,,\\
\Mp^2 \frac{\partial^2 N_e}{\partial \vp_*\partial\chi_*}
	&=&   \frac{\Mp}{W_*\sqrt{2\epsilon^*_\vp}}\frac{\partial\widetilde{Z}_e}{\partial \chi_*} ~~
	  =  -  \frac{\Mp}{W_*\sqrt{2\epsilon^*_\chi}}\frac{\partial\widetilde{Z}_e}{\partial \vp_*}\,,
\end{eqnarray}
where
$\bcalAs$ is defend as
\begin{eqnarray}
\bcalAs
	&\equiv&  \frac{W_e^2}{W_*^2}\frac{R\epsilon^e_\vp \epsilon^e_\chi}{\bepse^2}
		\left[\overline{\eta^e_{s}} - \left(R\epsilon^e_\vp + \frac{\epsilon^e_\chi}{R} \right)
			+\Mp\frac{ \epsilon^e_\vp \epsilon^e_\chi}{R\bepse}
			  \left\{ 		\frac{1}{\sqrt{2\epsilon^e_\vp}}\frac{\partial R}{\partial\vp_e}
				-  	\frac{R}{\sqrt{2\epsilon^e_\chi}}\frac{\partial R}{\partial \chi_e}
			\right\}
		\right] ,\quad \nonumber\\
\overline{\eta^e_{s}}& \equiv& \frac{R\epsilon^e_\vp \eta^e_{\chi\chi} + \epsilon^e_\chi \eta^e_{\vp\vp}}{\bepse} ~.
\end{eqnarray}
The derivatives of $\widetilde{Z}_e$ can be re-written as
\begin{eqnarray}
	   \sqrt{\epsilon^*_\vp} ~\frac{\partial \widetilde{Z}_e}{\partial \vp_*}
	=
-   \sqrt{\epsilon^*_\chi} ~\frac{\partial \widetilde{Z}_e}{\partial \chi_*}
	=
\frac{\sqrt{2}}{\Mp^2}W_* ~\bcalAs \,.
\end{eqnarray}

The contribution from the radiation dominated phase,  $\delta N_c$ can be evaluated from \eq{deltaNcSR}: 
\dis{
\delta N_c =\frac{(1-R)}{2\Mp}\frac{W_e}{W_*}\frac{\epsilon_\vp^e\epsilon_\chi^e}{\overline{\epsilon^e}}
	\left[   \frac{\delta \vp_*}{\sqrt{2\epsilon_\vp^*}}
		 -     \frac{\delta \chi_*}{\sqrt{2\epsilon_\chi^*}} \right].
}
We note that $\delta N_c$ has one more slow-roll parameter in the numerator than $\delta N_e$
and so suppressed.

In the case of $\delta N_e \gg \delta N_c$, the curvature perturbation $\zeta\simeq \delta N_e$ and
the power spectrum is
\begin{eqnarray}
\Ps_\zeta  &\simeq&
\frac{W_*}{24\pi^2 \Mp^4}\left(\frac{\baru^2}{\epsilon^*_\vp}+\frac{\barv^2}{\epsilon^*_\chi}\right)\,,
\label{eq:P} 
\end{eqnarray}
and spectral index and tensor-to-scalar ratio are
\begin{eqnarray}
n_\zeta-1  &\simeq &-2\epsilon^* -
		\frac{4}{\bu^2/\epsilon^*_\vp+\bv^2/\epsilon^*_\chi} \left[  \baru\left( 1-\frac{\eta^*_{\vp\vp}}{2\epsilon^*_\vp}~\baru\right) 
		          + \barv\left(1-\frac{\eta^*_{\chi\chi}}{2\epsilon^*_\chi}~\barv \right)\right]
 \, , \label{eq:n} \\
 r  &\simeq& \frac{2}{\pi^2 \Ps_\zeta} \frac{H_*^2}{\Mp^2} 
= \frac{16}{\bu^2/\epsilon^*_\vp+\bv^2/\epsilon^*_\chi}\, .\label{eq:r} 
\end{eqnarray}
The non-linear parameter is given by
\dis{
\frac{6}{5} \fnl  
 \simeq  \frac{2}{\left(\baru^2/\epsilon^*_\vp+\barv^2/\epsilon^*_\chi \right)^2} 
 \left[ \frac{\baru^2}{\epsilon^*_\vp}\left(1-\frac{\eta^*_{\vp\vp}}{2\epsilon^*_\vp}~\baru \right) 
		+ \frac{\barv^2}{\epsilon^*_\chi}\left(1-\frac{\eta^*_{\chi\chi}}{2\epsilon^*_\chi}~\barv \right) 
		+ \left(\frac{\baru}{\epsilon^*_\vp}-\frac{\barv}{\epsilon^*_\chi}\right)^2 \bcalAs \right]
	          \,.
  \label{fnlSum}
}
These obervables are again of the same form as in Ref.~\cite{VW} for the sum potential. However, we note again that the definitions of parameters are different and contains the effects from the end of inflation.

\subsection{Large non-Gaussianity generated during and at the  end of inflation}
In multiple scalar field inflation models, the large non-Gaussiniaty can be generated during inflation and the conditions for that has been studied by Byrnes, Choi and Hall~\cite{Byrnes:2008wi} with applications  to the multiple hybrid inflation in Ref.~\cite{Byrnes:2008zy}.
More general cases were studied, in the models beyond slow-roll in~\cite{Byrnes:2009qy} and 
in the general $n$-field inflation models in~\cite{Battefeld:2009ym}. Other studies on the non-Gaussianiy generated during inflation can be found in~\cite{Alabidi:2006hg,Kim:2010ud,Wang:2010si,Peterson:2010mv,Meyers:2010rg,Elliston:2011dr,Byrnes:2010em}.

Basically, the necessary condition for large non-Gaussianity to be generated during the inflation is  that one of the slow-roll parameters at the horizon exit is very small and it increases with time~\cite{Byrnes:2008wi}.  As discussed in~\cite{Byrnes:2008wi}, for inflation with two scalars $\vp$ and $\chi$,
 only if  $\epsilon^*_\vp\ll\epsilon^e_\vp\ll\epsilon^*_\chi \ll 1$ (or similarly  only if  $\epsilon^*_\chi\ll\epsilon^e_\chi\ll\epsilon^*_\vp \ll1$), the non-linearity parameter $\fnl$ 
can be large. That is to say, large non-Gaussianity is possible by a small slow-roll parameter $\epsilon_\vp$ or $\epsilon_\chi$ {\it at the horizon crossing}. 
For the product potential, for instance, this limit implies that $u \ll1$ and $v\simeq1$ so that $\eta_{\vp\vp^*}/\epsilon_\vp^*$  term dominate and ${\cal A}_p\simeq -u \, \eta_{\vp\vp}^e$. Accordingly,  
\dis{
\fnl \simeq \frac56 \frac{u^3}{(u^2 + \epsilon_\vp^*/\epsilon_\chi^*)^2} \left[ -\eta_{\vp\vp}^* +2 \eta_{\vp\vp}^e  \right].
} The sign of $\fnl$ is determined by the sign of $-\eta^*_{\vp\vp}+2\eta^e_{\vp\vp}$ 
for the case of $\epsilon^*_{\vp}\ll\epsilon^*_{\chi}$.   
However, this result is valid only when $R$ is the unity, i.e. the hypersurface of the end of  inflation is not very different from the uniform energy density hypersurface, as in the models of Refs.~\cite{Choi:2011me}. 

In a large class of multi-field inflation models, however, the hypersurface at the end of inflation can be quite different from that of the uniform energy density. In these cases  there could be another possibility for large non-Gaussianity: {\it the non-linearity parameter $\fnl$ can be large by the ending effect of inflation.} 

{\bf (a)}
From the similarity of the forms of $\fnl$  without the end effect given in the Ref.~\cite{Byrnes:2008wi} and  $\fnl$ with end effect in \eq{fnlPrd} and in \eq{fnlSum}, we can infer the conditions for large $\fnl$ in the case of $R\neq1$ including the end effect. Using the arguments in the previous paragraph, for small $\epsilon_\vp^*$, the condition for large $\fnl$ is $\epsilon^*_\vp\ll |R| \epsilon^e_\vp\ll \epsilon^*_\chi\ll 1$, therefore $\bu \ll 1$.
The new effect comes in with $R$ at the end of  inflation. Now there is no need for $\epsilon_\vp$ to increase upto the end of inflation. Instead, the condition $\epsilon^*_\vp\ll |R| \epsilon^e_\vp$ should be satisfied by a relatively large $|R|$. 
In this case, large non-linearity parameter approximately becomes
\dis{
\fnl \simeq \frac56 \frac{\baru^3}{(\baru^2 + \epsilon_\vp^*/\epsilon_\chi^*)^2} \left[ -\eta_{\vp\vp}^* +2 \eta_{\vp\vp}^e  \right], 
}
if $\p_\vp R$ and $\p_\chi R$ are suppressed enough. 
The similar expression holds also for small $\epsilon_\chi^*$ with small $|R|$. In this case, the required condition is $\epsilon_\chi^* \ll |R|^{-1}\epsilon_\chi^e \ll \epsilon_\vp^* \ll 1$.

{\bf (b)}
In the limit of $R\rightarrow\infty$ (or $0$), 
$\bar{u}$ (or $\bar{v}$) approach to the unity, while $\bar{v}$ (or $\bar{u}$) vanish. If $\p_\vp R$ and $\p_\chi R$ are not divergent and {\it $\epsilon^*_\vp$ and $\epsilon^*_\chi$ are not hierarchical}, $\fnl$ becomes just of order $\epsilon^*_{\vp}$ or $\epsilon^*_{\chi}$ in this limit, which is quite suppressed. 
However, a large $\p_\chi R$ (or $\p_\vp R$) can result in a large $\fnl$, when $\epsilon_\chi\simeq\eta_{\chi\chi}\simeq 0$, thus  $R=\sqrt{\epsilon_\chi^e/\epsilon_\vp^e} (E_\vp/E_\chi)\simeq 0$
in \eq{fnlPrd},
\dis{ \label{endfnl}
\frac65\fnl\simeq -\frac{2}{(1+ \bu^2  \epsilon_\chi^* /\bv^2 \epsilon_\vp^*)^2}\frac{\bcalAp}{\bv^2} \simeq 
\frac{1}{(1+ \bu^2  \epsilon_\chi^* /\bv^2 \epsilon_\vp^*)^2} \frac{2\epsilon_\vp}{R}\left( \frac{1}{\sqrt{2\epsilon_\vp}} \frac{\p R}{\p \vp} - \frac{R}{\sqrt{2\epsilon_\chi}}\frac{\p R}{\p \chi} \right)
\,,}
for $\Mp = 1$.
 In this case, 
{\it a large $\fnl$ is generated only by the end effect of inflation}. We will discuss a specific example later.

{\bf (c)}
As pointed earlier, $R$ can be negative and so $\overline{\epsilon^e}$ has a vanishing limit. It opens a new possibility for large non-Gaussianity. 
Consider the case that $\overline{\epsilon^e}$ in Eqs.~(\ref{R}) and (\ref{Rsum}) is extremely small, 
\dis{ \label{condi}
\overline{\epsilon^e}=R\epsilon^e_\vp+\epsilon^e_\chi
\longrightarrow\varepsilon ~, 
} 
where $\varepsilon$ is assumed to be very small 
($\ll\epsilon^*_{\vp,\chi},~\epsilon^e_{\vp,\chi}$).  
Note that the parameters appearing in the definition of $\overline{\epsilon^e}$ are all associated with the field values at the end of inflation. Under the limit of $\overline{\epsilon^e}\rightarrow\varepsilon$ or $\epsilon_\chi^e \simeq -R \epsilon_\vp^e$, $\bar{u}$, $\bar{v}$ in \eq{uvbar} and $\bu$, $\bv$, $\widetilde{Z}_e$ in Eqs.~(\ref{uv}), (\ref{Ztilde}) diverge as ${\cal O}(1/\varepsilon)$. Since $\bcalAp$ and $\bcalAs$ are divergent as ${\cal O}(1/\varepsilon^3)$ in this limit, the third terms in Eqs.~(\ref{fnlPrd}) and (\ref{fnlSum}) become dominant in $\fnl$, and so the large non-linearity parameter $\fnl$ becomes also divergent as ${\cal O}(1/\varepsilon)$. 
 In this case $\fnl$ is approximately 
\dis{
\fnl\simeq -\frac56\frac{\epsilon_\chi^e}{\varepsilon}\frac{2(\epsilon_\chi^*-\epsilon_\vp^*)^2}{(\epsilon^*)^2} \left[ \eta^e_\vp - \eta^e_\chi +2\left(1+\frac{1}{R}\right)\epsilon_\chi^e- \frac{\sqrt{\epsilon_\chi^e}}{R^2\sqrt2 } \left(\sqrt{-R} \frac{\p R}{\p \vp} -R \frac{\p R}{\p \chi} \right) \right]\,,}
setting $\Mp = 1$.

\section{Application}
In this section we apply our results obtained in the previous section to the several specific models.
\label{application}

\subsection{Generating curvature perturbation at the end of inflation}
As a simple example, let us discuss a scenario in which  one field $\phi$ drives the inflation, while the other $\sigma$ is negligible during inflation, since its potential is so flat to make practically no effect on the inflation  trajectory.  During inflation the potential is, hence,  $W(\phi,\sigma)\simeq U(\phi)$. More specific potentials are given in~\cite{Lyth:2005qk,Lyth:2006nx,Alabidi:2006hg}.  

Inflation ends when 
inflaton field $\phi$ has a certain value $\phi_e$. However the $\phi_e$ depends on the field
$\sigma$. Since $\phi_e(\sigma)$ would depend on the position through the perturbation $\delta\sigma(\bf x)$, curvature perturbation should have an additional contribution $\zeta_e$ coming from the end effect as well as the pre-existing perturbation by the inflaton field $\zeta_{\phi}$.

The $\sigma$ field does not move during inflation, $\sigma_e\simeq\sigma_*$, and $\epsilon_\sigma \rightarrow 0$ since its potential is so flat.
As the condition of the fields at the end of inflation, we use $E(\phi_e,\sigma_e)=E(\phi_e,\sigma_*)=$ a constant from \eq{endcondition}.

For this case, \eq{dervsproduct} leads to
\dis{
\frac{\p \phi_e}{\p \phi_*}=\frac{\p \sigma_e}{\p \phi_*}=0,\quad
\frac{\p \phi_e}{\p \sigma_*}=-\frac{E_\sigma}{E_\phi} , \quad\text{and}\quad \frac{\p \sigma_e}{\p \sigma_*}=1,
}
and
\dis{
\bu =1,\qquad \bv =\frac{\epsilon_\sigma^e}{R\epsilon_\phi^e},\qquad
 \text{with} \qquad 
R=\sqrt{\frac{\epsilon_\sigma^e}{\epsilon_\phi^e}} \frac{E_\phi}{E_\sigma}.
\label{R-Erel}
}
Here we can see that $\bv \rightarrow 0$ and $R\rightarrow 0 $.
It is easy to find the first derivatives of e-folding number, 
\dis{
\frac{\p N_e}{\p \phi_*} = \frac{1}{\Mp \sqrt{2\epsilon_\phi^*}},
\qquad \frac{\p N_e}{\p \sigma_*} = \frac{1}{\Mp \sqrt{2\epsilon_\phi^e}} \frac{E_\sigma}{E_\phi},\label{linear_1}
}
where the small value of $\epsilon_\sigma$ was canceled and disappeared.
The second derivatives are given as
\dis{
\Mp^2\frac{\p^2 N_e}{\p \phi_{*}^2}&=\left( 1-\frac{\eta_{\phi\phi}^*}{2\epsilon^*_\phi} \right),\\
\Mp^2\frac{\p^2 N_e}{\p \sigma_*^2}&=-\left( 1-\frac{\eta_{\phi\phi}^e}{2\epsilon_{\phi}^e} \right)\bfrac{E_\sigma}{E_\phi}^2 + 
 \frac{\Mp}{ \sqrt{2\epsilon_\phi^e}} \frac{\p }{\p \sigma_*}\bfrac{E_\sigma}{E_\phi},\\
 \Mp^2\frac{\p^2 N_e}{\p \sigma_*\p\phi_*}&= \Mp^2\frac{\p^2 N}{\p \phi_*\p\sigma_*}=0.\label{second_1}
}

The curvature perturbation $\delta N_e$ can be determined by the two contributions: one is $\zeta_{\phi}$ coming from the perturbation of inflaton field $\delta \phi$ and the other is $\zeta_\sigma$ from $\delta \sigma$, i.e.
\dis{
\zeta\simeq \delta N_e = \zeta_\phi + \zeta_\sigma,
}
where
\dis{
\zeta_\phi &= \frac{\p N_e}{\p \phi_*}  \delta \phi_* +\frac12 \frac{\p^2 N_e}{\p \phi_*^2}(\delta \phi_* )^2,\\
\zeta_\sigma &= \frac{\p N_e}{\p \sigma_*} \delta\sigma_* +\frac12 \frac{\p^2 N_e}{\p \sigma_*^2} (\delta\sigma_*)^2.
}
Here the first and second derivatives are given in Eqs.~(\ref{linear_1}) and~(\ref{second_1}).

Then the Power spectrum is given by
\dis{
\calP_\zeta&= \frac{1}{2\Mp^2\epsilon_\phi^*}\left[1+ \frac{\epsilon_\phi^*}{\epsilon_\phi^e} \bfrac{E_\sigma}{E_\phi}^2  \right] \bfrac{H_*}{2\pi}^2.
}
 $\zeta_\sigma$ can dominate $\zeta_\phi$, when 
\dis{
\frac{E_\sigma^2}{ E_\phi^2} \gg \frac{\epsilon_\phi^e}{\epsilon_\phi^*}.
}
For  $\zeta_\sigma\gg\zeta_\phi$, the non-linearity parameter $\fnl$ becomes 
 \dis{
\frac65\fnl\simeq \left. \frac{\p^2 N_e}{\p \sigma_*^2} \right/ \left(   \frac{\p N_e}{\p \sigma_*} \right)^2 
= -(2\epsilon_\phi^e-\eta_\phi^e) + \Mp \sqrt{2 \epsilon_\phi^e} \bfrac{E_\sigma}{E_\phi}^{-2}\frac{\p}{\p\sigma_*} \bfrac{E_\sigma}{E_\phi}.
}
Thus, a large $\fnl$ is possible when the second term is large. Note that it can be obtained from \eq{endfnl} using \eq{R-Erel}. Accordingly, it is classified to ``{\bf (b)}'' discussed in the previous section.

%%%%%%%%%%%%%%%%%%%%%%%%%%%%%%%%%%%%%%%%%%%%%%%%%%%%%
%%%%%%%%%%%%%%%%%%%%%%%%%%%%%%%%%%%%%%%%%%%%%%%%%%%%%

\subsection{Hybrid inflation with general ending}

In this subsection we consider the hybrid-type inflation with two scalar fields $\phi_1$ and $\phi_2$ during inflation in addition to a waterfall field $\psi$.
The hybrid inflation can be realized with the potential 
\dis{
W=W_0(\phi_1,\phi_2,\psi)W_{\rm inf}(\phi_1,\phi_2)~,
}
where $W_0$ is given by
\begin{eqnarray}
&&W_0=\frac12 G(\phi_1,\phi_2) \psi^2 +\frac{\lambda}{4} \left( \psi^2-\frac{\sigma^2}{\lambda} \right)^2 ~.
\end{eqnarray}
Here  a mass term for $\psi$, $G(\phi_1,\phi_2)$, can take the form~\cite{Naruko:2008sq}, 
\dis{
G(\phi_1,\phi_2) \equiv g_1^2 (\phi_1\cos\alpha +\phi_2\sin\alpha)^2 +g_2^2 (-\phi_1\sin\alpha + \phi_2\cos\alpha)^2 ~,\label{Gend}
}
where $g_1$ and $g_2$ are dimensionless constants.
During inflation  when $G(\phi_1,\phi_2)$ is large enough compared to $\sigma^2$,
 the waterfall field $\psi$ is heavy and  stuck to the origin, inflating the universe with the positive vacuum energy $W_0=\sigma^4/4\lambda$. 
When $G$ becomes smaller than $\sigma^2$, however,  $\psi$ rolls down to the true minimum ($\psi^2=\sigma^2/\lambda$), terminating inflation. 
Hence, $G(\phi_1,\phi_2)$ yields the condition for the end of inflation and can be identified
as \eq{endcondition} introduced in the previous section. Identifying $E=G$, we find that
\dis{
E_1 &\equiv \frac{\p E}{\p \phi_1}= 2g_1^2 (\phi_1\cos\alpha +\phi_2\sin\alpha) \cos\alpha - 2g_2^2(-\phi_1\sin\alpha + \phi_2\cos\alpha)\sin\alpha ~, \\
E_2&\equiv \frac{\p E}{\p \phi_2}= 2g_1^2 (\phi_1\cos\alpha +\phi_2\sin\alpha) \sin\alpha +2g_2^2(-\phi_1\sin\alpha + \phi_2\cos\alpha)\cos\alpha ~.
}

For a clear discussion, we consider the specific potential for $W_{\rm inf}$ used in Refs.~\cite{Naruko:2008sq,Alabidi:2006hg,Byrnes:2008zy}, given by   
\dis{
W_{\rm inf}(\phi_1,\phi_2) = \exp\left[\frac12 m_1^2\phi_1^2+\frac12m_2^2\phi_2^2\right] .
}
For simplicity, throughout this subsection we set $M_P=1$ and drop the super- (sub-) scripts ``$e$'' indicating end of inflation. 

Particularly, in the case of the equal mass $m_1=m_2\equiv m$, one take  $\alpha=0$  in the condition of the end of inflation, because of the rotational symmetry. Thus, $W_{\rm inf}=\exp\left[ m^2(\phi_1^2+\phi_2^2)/2  \right]$ and $E=g_1^2\phi_1^2 + g_2^2\phi_2^2$. 
With this potential, the slow-roll parameters and  $R$ defined in \eq{R} becomes 
\dis{ \label{SRend}
\epsilon_{1}=\frac{1}{2}m^4\phi_1^2 ~,&\qquad\epsilon_{2}=\frac{1}{2}m^4\phi_2^2 ~,\\
\eta_{1}= m^2(1+m^2\phi_1^2) ~,&\qquad\eta_{2}= m^2(1+m^2\phi_2^2) ~.
}
From  $E_1= 2g_1^2 \phi_1$ and $E_2= 2g_2^2 \phi_2$ , we have 
\dis{
R=\frac{m_2^2\phi_2}{m_1^2\phi_1}\, \bfrac{g_1^2\phi_{_1}}{g_2^2\phi_{_2}}= \frac{ g_1^2}{ g_2^2}, 
}
which is independent of $\phi_1$ and $\phi_2$.
Hence,  
\dis{ \label{epsilonbar}
\beps =R\epsilon_1+\epsilon_2 = \frac{m^4\sigma^2}{2g_2^2},
}
where $\sigma^2=g_1^2\phi_1^2 + g_2^2\phi_2^2 $ is a constant.

For $m_1=m_2$ ($=m$) and $\alpha=0$, thus, the power spectrum is given by 
\dis{
\calP_\zeta = \frac{1}{2} \frac{E_1^2 \epsilon_1/\epsilon_1^* + E_2^2 \epsilon_2/\epsilon_2^*}{( E_1\sqrt{\epsilon_1} +E_2\sqrt{\epsilon_2}  )^2}  \bfrac{H_*}{2\pi}^2 = \frac{e^{-2m^2N}}{m^4\sigma^2} (g_1^4\phi_1^2 +g_2^4 \phi_2^2) \bfrac{H_*}{2\pi}^2   ,
}
and the non-linear parameter $\fnl$ is
\dis{
\frac65 \fnl= \frac{4m^2}{g_1^4\phi_1^2 +g_2^4 \phi_2^2} \left[ - \sigma^2(g_1^6\phi_1^2 +g_2^6\phi_2^2) +2(g_1^2-g_2^2) g_1^2g_2^2\phi_1^2\phi_2^2\right].
}

\section{Conclusion}                                           
\label{conclusion}                                          
In this paper, we have obtained a formula for curvature perturbation in multi-field inflation models. The formula contains the contributions from the evolution of perturbation during inflation and also from the effect at the end of inflation. For clearness of discussion, we utilized the potentials separable by product or sum with two scalar fields, but it is manifest to extend our discussion to cases of more scalar fields.
Using the derived expression for curvature perturbation, we could get the analytic formulae for the power spectrum, spectral index and non-linearity parameter. Especially we could find more possibilities for large non-Gaussianity $\fnl$ generated due to the effects by the end of inflation in addition to the generation during inflation. This formalism can be extended to accommodate more scenarios for the generation of curvature perturbation e.g. by the curvaton or modulated reheating~\cite{Alabidi:2010ba}.

\acknowledgments{ \noindent  
This work was supported by the 
National Research Foundation of Korea (NRF) grant funded by the 
Korea government (MEST) (No. 2010-0009021 and No. 2011-0011083).
K.Y.C. and S.A.Kim acknowledge the Max Planck Society (MPG), the Korea Ministry of
Education, Science and Technology (MEST), Gyeongsangbuk-Do and Pohang
City for the support of the Independent Junior Research Group at the Asia Pacific
Center for Theoretical Physics (APCTP).
}

\section{Appendix}

In this Appendix we show that our formulae give the same results with those in two-brid inflation studied in  Ref.~\cite{Naruko:2008sq}.  
The relevant potential is 
\dis{
W=W_0(\phi_1,\phi_2,\psi)W_{\rm inf}(\phi_1,\phi_2)~,
}
where $W_0$ and $W_{\rm inf}$ are given by
\begin{eqnarray}
&&W_0=\frac12 G(\phi_1,\phi_2) \psi^2 +\frac{\lambda}{4} \left( \psi^2-\frac{\sigma^2}{\lambda} \right)^2 ~.
\end{eqnarray}
\dis{
W_{\rm inf}(\phi_1,\phi_2) = \exp\left[\frac12 m_1^2\phi_1^2+\frac12m_2^2\phi_2^2\right] .
}
For simplicity, we set $M_P=1$ and drop the super-(and  sub-)scripts ``$e$'' indicating end of inflation.
Here the mass term for $\psi$, $G(\phi_1,\phi_2)$ can take the form~\cite{Naruko:2008sq},
\dis{
G(\phi_1,\phi_2) \equiv g_1^2 (\phi_1\cos\alpha +\phi_2\sin\alpha)^2 +g_2^2 (-\phi_1\sin\alpha + \phi_2\cos\alpha)^2 ~,\label{Gend}
}
where $g_1$ and $g_2$ are dimensionless constants.
The condition for ending inflation is $E(\phi_1,\phi_2)=G(\phi_1,\phi_2)=\sigma^2$.

For comparison, we use the same notation $X,Y,W$ and $Z$ defined in Ref.~\cite{Naruko:2008sq}, 
\dis{ \label{phi_12}
&\phi_1=\frac{\sigma}{g_1g_2}\left(g_2{\rm cos}\alpha{\rm cos}\gamma -g_1{\rm sin}\alpha{\rm sin}\gamma\right)\equiv \left(\frac{g\sigma}{g_1g_2}\right)X ~,
\\
&\phi_2=\frac{\sigma}{g_1g_2}\left(g_2{\rm sin}\alpha{\rm cos}\gamma +g_1{\rm cos}\alpha{\rm sin}\gamma\right)\equiv \left(\frac{g\sigma}{g_1g_2}\right)Z ~,
\\
&\quad\quad\quad \frac{\p}{\p\gamma}X\equiv -Y ~,\quad{\rm and}\quad \frac{\p}{\p\gamma}Z\equiv -W ~. 
}
They satisfy
\dis{ \label{rel}
 \frac{X}{Z} =\frac{\phi_{1}}{\phi_{2}} ~,\quad \frac{W}{Y}=- \frac{E_1}{E_2} ~,\quad
 R= -\frac{m_2^2}{m_1^2}\frac{WZ}{XY}=-\frac{m_2^2\phi_2^2}{m_1^2\phi_1^2}\frac{XW}{YZ} ~,
}
and 
\dis{ \label{epsilonbar}
\beps =R\epsilon_1+\epsilon_2 = \frac{1}{2} m_2^2\phi_2^2 \bfrac{X}{Y} \left(m_2^2 \frac{Y}{X} - m_1^2 \frac{W}{Z}  \right).
}
Other useful relations are
\dis{ \label{relations}
&\qquad\qquad \frac{\p}{\p\phi_{1}} \bfrac{E_1}{E_2} = \frac{g_1g_2}{\sigma g}\frac{Z}{Y^2} ~, \quad
\frac{\p}{\p\phi_{2}} \bfrac{E_1}{E_2} = -\frac{g_1g_2}{\sigma g}\frac{X}{Y^2} ~,\\
&\qquad \frac{\p}{\p\phi_{1}} \bfrac{E_1}{E_2}-   \bfrac{E_1}{E_2}\frac{\p}{\p\phi_{2}} \bfrac{E_1}{E_2} =-\frac{g_1g_2}{\sigma g} \bfrac{ZX}{Y^3} \left(\frac{W}{Z} -\frac{Y}{X}\right) ~,
\\
&\frac{\p R}{\p\phi_{1}}=\frac{m_2^2}{m_1^2}\frac{g_1g_2}{g\sigma}\frac{Z}{XY}\left(\frac{W}{X}+\frac{Z}{Y}\right)~,\quad{\rm and}\quad \frac{\p R}{\p\phi_{2}}=-\frac{m_2^2}{m_1^2}\frac{g_1g_2}{g\sigma}\frac{Z}{XY}\left(\frac{W}{Z}+\frac{X}{Y}\right)~.
}

Using the equation of motion of the fields in the slow-roll limit, the field values at the horizon exit, $\phi_1^*$ and $\phi_2^*$ can be estimated by tracing back the field values at the end of inflation, $\phi_1$ and $\phi_2$ so as to achieve 55-65 e-folds:
\dis{
N_e\approx\int^*_e\frac{d\phi_{i}}{m_{i}^2~\phi_{i}}=\frac{1}{m_{i}^2}~{\rm log}\frac{\phi^*_{i}}{\phi_{i}} ~,
}
where $i=$ 1 or 2.
The slow-roll parameters at the horizon crossing, $\epsilon^*_{1}$,  $\epsilon^*_{2}$ and $\eta^*_{1}$, $\eta^*_{2}$ are given by 
\dis{  \label{inlVal}
&\qquad\qquad\epsilon^*_{i}=\frac12 m_{i}^4\phi^{*2}_{i}=\frac12 m^4_{i}\phi_{i}^2~e^{2N_em^2_{i}} ~,
\\
&\eta^*_{i}
=m^2_{i}\left(1+m^2_{i}\phi^{*2}_{i}\right)
=m^2_{i}\left(1+m^2_{i}\phi^{2}_{i}
~e^{2N_em^2_{i}}\right) .
}

After some manipulations with the above relations, the $\delta N_e$ given by \eq{linearproduct} and \eq{secondproduct} is easily converted to give: 
\dis{
\delta N_e=&  \sum_I
N_{e,I}\delta\phi_{I*}+\frac12\sum_{IJ}N_{e,IJ}\delta\phi_{I*}\delta\phi_{J*},
\\
=& \frac{\displaystyle{-\frac{W}{Z} \frac{\delta \phi_1^*}{\phi_1^*}+\frac{Y}{X}  \frac{\delta \phi_2^*}{\phi_2^*}   }}{\displaystyle{m_2^2 \frac{Y}{X}- m_1^2 \frac{W}{Z} }}
  +  \frac12\frac{\displaystyle{\frac{W}{Z} \bfrac{\delta \phi_1^*}{\phi_1^*}^2-\frac{Y}{X}  \bfrac{\delta \phi_2^*}{\phi_2^*}^2   }}{\displaystyle{m_2^2 \frac{Y}{X}- m_1^2 \frac{W}{Z} } }
\\
&  -\frac12\frac{\displaystyle{\left( 1 -\frac{Y}{X}\frac{W}{Z} \right) \left(\frac{W}{Z} -\frac{Y}{X}\right)\left(\frac{m_2^2}{\phi_1^*}\delta\phi_1^*  -\frac{m_1^2}{\phi_2^*}\delta\phi_2^* \right)^2   }}{\displaystyle{\left(m_2^2 \frac{Y}{X}- m_1^2 \frac{W}{Z} \right)^3}} +\cdots.
}
which is exactly the same as the result of  Ref.~\cite{Naruko:2008sq}.

More explicitly we can check out our formulae for the power spectrum and non-linear parameter.
With the definitions of $\overline{u}$ and $\overline{v}$ of \eq{uvbar}, the initial values of the slow-roll parameters of \eq{inlVal}, and Eqs.~(\ref{SRend}), (\ref{epsilonbar}), (\ref{phi_12}), (\ref{rel}), we get  
\dis{ \label{denorm}
\frac{\overline{u}^2}{\epsilon^*_{1}}+\frac{\overline{v}^2}{\epsilon^*_{2}}=&\frac{m_2^4}{2\overline{\epsilon}^2}\left(\frac{g\sigma}{g_1g_2}\right)^2\frac{Z^2}{Y^2}\left[W^2e^{-2N_em_1^2}+Y^2e^{-2N_em_2^2}\right]  \\
=&2\left(\frac{g_1g_2}{g\sigma}\right)^{2}
\frac{\left[W^2e^{-2N_em_1^2}+Y^2e^{-2N_em_2^2}\right]}
{\left(m_2^2YZ-m_1^2XW\right)^2} ~.
}
This relation recasts the expression of the power spectrum given in \eq{powerPrd} to 
\begin{equation}
{\cal P}_\zeta = \frac12\left(\frac{\overline{u}^2}{\epsilon^*_{1}}+\frac{\overline{v}^2}{\epsilon^*_{2}}\right)\left(\frac{H_*}{2\pi}\right)^2
=\left(\frac{g{\rm sin}\beta{\rm cos}\beta}{\sigma}\right)^2
\frac{\left[W^2e^{-2N_em_1^2}+Y^2e^{-2N_em_2^2}\right]}
{\left(m_2^2YZ-m_1^2XW\right)^2}\left(\frac{H_*}{2\pi}\right)^2 ,
\end{equation}
where ${\rm cos}\beta\equiv g_1/g$ and ${\rm sin}\beta\equiv g_2/g$. 
This is again exactly coincident with the expression of Ref.~\cite{Naruko:2008sq}.  
Similarly, one can confirm also the spectral index $n_\zeta$ and the scalar to tensor ratio.

Using Eqs.~(\ref{SRend}), (\ref{phi_12}), (\ref{rel}), and (\ref{inlVal}), 
the first two terms of the numerator in \eq{fnlPrd} also can be evaluated in terms of the parameters employed in Ref.~\cite{Naruko:2008sq} as follows;
\begin{eqnarray} \label{1st+2nd}
&&\frac{\overline{u}^3}{\epsilon^*_{1}}\left(1-\frac{\eta^*_{1}}{2\epsilon^*_{1}}\right)+\frac{\overline{v}^3}{\epsilon^*_{2}}\left(1-\frac{\eta^*_{2}}{2\epsilon^*_{2}}\right)=\frac{m_2^6}{4\overline{\epsilon}^3}\left(\frac{g\sigma}{g_1g_2}\right)^2\frac{Z^3}{Y^3}\left[\frac{W^3}{X}e^{-4N_em_1^2}-\frac{Y^3}{Z}e^{-4N_em_2^2}\right] .~~~~~
\end{eqnarray}
Now let us calculate the third term of \eq{fnlPrd} piece by piece. 
With Eqs.~(\ref{SRend}), (\ref{phi_12}), (\ref{rel}),  (\ref{relations}), and (\ref{inlVal}), we get the following piece expressions for calculation of the third term of $\fnl$ in \eq{fnlPrd};
\begin{eqnarray}
&&\left(\frac{\overline{u}}{\epsilon^*_{1}}
-\frac{\overline{v}}{\epsilon^*_{2}}\right)^2
=\frac{1}{\overline{\epsilon}^2m_1^4}
\frac{Z^2}{Y^2}\left(m_2^2\frac{W}{X}e^{-2N_em_1^2}+m_1^2\frac{Y}{Z}e^{-2N_2m_2^2}\right)^2 , \\
&&-\frac{\epsilon_{1}\epsilon_{2}}{\overline{\epsilon}^3}R=\frac{m_1^2m_2^6}{4\overline{\epsilon}^3}
\left(\frac{g\sigma}{g_1g_2}\right)^4\frac{XWZ^3}{Y} ~, \nonumber \\
&&\epsilon_{2}\eta_{1}+R\epsilon_{1}\eta_{2}
=\frac{m_1^2m_2^4}{2}\left(\frac{g\sigma}{g_1g_2}\right)^2\left[Z^2-\frac{XWZ}{Y}+m_1^2\left(\frac{g\sigma}{g_1g_2}\right)^2X^2Z^2\left(1-\frac{m_2^2}{m_1^2}\frac{WZ}{XY}\right)\right] , \nonumber \\
&&-2(1+R)\epsilon_{1}\epsilon_{2}=-\frac{m_1^4m_2^4}{2}\left(\frac{g\sigma}{g_1g_2}\right)^4X^2Z^2\left(1-\frac{m_2^2}{m_1^2}\frac{WZ}{XY}\right) , \nonumber \\
&&\frac{1}{R}(\partial_{\phi_1}R)\epsilon_{2}
\sqrt{\frac{\epsilon_{1}}{2}}
-(\partial_{\phi_2}R)\epsilon_{1}
\sqrt{\frac{\epsilon_{2}}{2}}=\frac{m_1^2m_2^4}{4}\left(\frac{g\sigma}{g_1g_2}\right)^2\left[-Z^2+\frac{XWZ}{Y}-\frac{XZ^3}{YW}+\frac{X^2Z^2}{Y^2}\right] . \nonumber 
\end{eqnarray}
Assembling the above small fragments gives the full expression for the third term of the numerator in \eq{fnlPrd}:
\begin{eqnarray} \label{3rd}
\left(\frac{\overline{u}}{\epsilon^*_{1}}
-\frac{\overline{v}}{\epsilon^*_{2}}\right)^2 \bcalAp = 
\frac{m_2^6}{4\overline{\epsilon}^3}\left(\frac{g\sigma}{g_1g_2}\right)^2\frac{Z^3}{Y^3}\left[\frac{(XZ-YW)\left(\frac{W}{Z}-\frac{Y}{X}\right)\left(m_2^2\frac{W}{X}e^{-2N_em_1^2}+m_1^2\frac{Y}{Z}e^{-2N_em_2^2}\right)^2}{\left(m_2^2\frac{Y}{X}-m_1^2\frac{W}{Z}\right)^2}\right] . \nonumber \\
\end{eqnarray}
By inserting Eqs.~(\ref{denorm}), (\ref{1st+2nd}), (\ref{3rd}), and (\ref{epsilonbar}), (\ref{phi_12}) into Eq.~(\ref{fnlPrd}), we can reproduce the result 
of Ref.~\cite{Naruko:2008sq}:
\begin{eqnarray}
\frac65 f_{\rm NL}&=&\frac{XZ}{\left(m_2^2\frac{Y}{X}-m_1^2\frac{W}{Z}\right)
\left(Y^2e^{2N_em_1^2}+W^2e^{2N_em_2^2}\right)^2}
\Bigg[
\left(\frac{W^3}{X}e^{4N_em_2^2}
-\frac{Y^3}{Z}e^{4N_em_1^2}\right)
\left(m_2^2\frac{Y}{X}-m_1^2\frac{W}{Z}\right)^2
\nonumber \\
&&\quad\quad\quad -\left(XZ-YW\right)\left(\frac{W}{Z}-\frac{Y}{X}\right)
\left(m_1^2\frac{Y}{Z}e^{2N_em_1^2}
+m_2^2\frac{W}{X}e^{2N_em_2^2}\right)^2\Bigg] ~.
\label{fnlquad}
\end{eqnarray}
In Ref.~\cite{Naruko:2008sq}, the authors discussed the possibility of large non-Gaussianity based on the above expression \eq{fnlquad} for the cases that the two inflaton masses are degenrate ($m_1^2=m_2^2$) and hierarchical ($m_1^2\gg m_2^2$).

%%%%%%%%%%%%%%%%%%%%%%%%%%%%%%%%%%%%%%%%%%%%%%%%%%%%%%%%%


\begin{thebibliography}{99}


\bibitem{Komatsu:2010fb}
  E.~Komatsu {\it et al.},
  %``Seven-Year Wilkinson Microwave Anisotropy Probe (WMAP) Observations:
  %Cosmological Interpretation,''
  arXiv:1001.4538 [astro-ph.CO].


%\cite{Bassett:2005xm}
\bibitem{Bassett:2005xm}
  B.~A.~Bassett, S.~Tsujikawa and D.~Wands,
  %``Inflation dynamics and reheating,''
  Rev.\ Mod.\ Phys.\  {\bf 78} (2006) 537
  [astro-ph/0507632].
  %%CITATION = ASTRO-PH/0507632;%%
%\cite{Wands:2007bd}
\bibitem{Wands:2007bd}
  D.~Wands,
  %``Multiple field inflation,''
  Lect.\ Notes Phys.\  {\bf 738} (2008) 275
  [astro-ph/0702187 [ASTRO-PH]].
  %%CITATION = ASTRO-PH/0702187;%%
  
  
%\cite{Byrnes:2008wi}
\bibitem{Byrnes:2008wi}
  C.~T.~Byrnes, K.~Y.~Choi and L.~M.~H.~Hall,
  %``Conditions for large non-Gaussianity in two-field slow-roll inflation,''
  JCAP {\bf 0810} (2008) 008
  [arXiv:0807.1101 [astro-ph]].
  %%CITATION = JCAPA,0810,008;%%

  
%\cite{Lyth:2005qk}
\bibitem{Lyth:2005qk}
  D.~H.~Lyth,
  %``Generating the curvature perturbation at the end of inflation,''
  JCAP {\bf 0511 } (2005)  006.
  [astro-ph/0510443].
  
  %\cite{Lyth:2006nx}
\bibitem{Lyth:2006nx}
  D.~H.~Lyth, A.~Riotto,
  %``Generating the Curvature Perturbation at the End of Inflation in String Theory,''
  Phys.\ Rev.\ Lett.\  {\bf 97 } (2006)  121301.
  [astro-ph/0607326].
 
  
%\cite{Sasaki:2008uc}
\bibitem{Sasaki:2008uc}
  M.~Sasaki,
  %``Multi-brid inflation and non-Gaussianity,''
  Prog.\ Theor.\ Phys.\  {\bf 120 } (2008)  159-174.
  [arXiv:0805.0974 [astro-ph]].

%\cite{Naruko:2008sq}
\bibitem{Naruko:2008sq}
  A.~Naruko, M.~Sasaki,
  %``Large non-Gaussianity from multi-brid inflation,''
  Prog.\ Theor.\ Phys.\  {\bf 121 } (2009)  193-210.
  [arXiv:0807.0180 [astro-ph]].

%\cite{Huang:2009xa}
\bibitem{Huang:2009xa}
  Q.~-G.~Huang,
  %``The Trispectrum in the Multi-brid Inflation,''
  JCAP {\bf 0905} (2009) 005
  [arXiv:0903.1542 [hep-th]].
  %%CITATION = ARXIV:0903.1542;%%   

  
    
  %\cite{Byrnes:2008zy}
\bibitem{Byrnes:2008zy}
  C.~T.~Byrnes, K.~-Y.~Choi, L.~M.~H.~Hall,
  %``Large non-Gaussianity from two-component hybrid inflation,''
  JCAP {\bf 0902 } (2009)  017.
  [arXiv:0812.0807 [astro-ph]].
  
%\cite{Huang:2009vk}
\bibitem{Huang:2009vk}
  Q.~-G.~Huang,
  %``A Geometric description of the non-Gaussianity generated at the end of multi-field inflation,''
  JCAP {\bf 0906} (2009) 035
  [arXiv:0904.2649 [hep-th]].
  %%CITATION = ARXIV:0904.2649;%%

  
%\cite{Yokoyama:2008xw}
\bibitem{Yokoyama:2008xw}
  S.~Yokoyama and J.~Soda,
  %``Primordial statistical anisotropy generated at the end of inflation,''
  JCAP {\bf 0808} (2008) 005
  [arXiv:0805.4265 [astro-ph]].
  %%CITATION = ARXIV:0805.4265;%%

%\cite{Emami:2011yi}
\bibitem{Emami:2011yi}
  R.~Emami and H.~Firouzjahi,
  %``Issues on Generating Primordial Anisotropies at the End of Inflation,''
  arXiv:1111.1919 [astro-ph.CO].
  %%CITATION = ARXIV:1111.1919;%%
  
  
\bibitem{starob85}
 A.~A.~Starobinsky,
%  {\em Multicomponent De Sitter (Inflationary) Stages And The Generation Of  Perturbations,}
  JETP Lett.\  {\bf 42}, 152 (1985)
  [Pisma Zh.\ Eksp.\ Teor.\ Fiz.\  {\bf 42}, 124 (1985)].

\bibitem{ss1}
 M.~Sasaki   and E.~D.~Stewart,
%  {\em A General analytic formula for the spectral index of the density
%  perturbations produced during inflation,}
  Prog.\ Theor.\ Phys.\  {\bf 95} (1996) 71
[arXiv:astro-ph/9507001].

\bibitem{Sasaki:1998ug}
 M.~Sasaki and T.~Tanaka,
% {\em Super-horizon scale dynamics of multi-scalar inflation,}
 Prog.\ Theor.\ Phys.\  {\bf 99}, 763 (1998)
 [arXiv:gr-qc/9801017].






\bibitem{lms}
D.~H.~Lyth, K.~A.~Malik and M.~Sasaki,
%{\em A general proof of the conservation of the curvature perturbation,}
JCAP {\bf 0505}, 004 (2005)
[arXiv:astro-ph/0411220].


  
%\cite{Maldacena:2002vr}
\bibitem{Maldacena:2002vr}
  J.~M.~Maldacena,
  %``Non-Gaussian features of primordial fluctuations in single field
  %inflationary models,''
  JHEP {\bf 0305}, 013 (2003)
  [arXiv:astro-ph/0210603].
  %%CITATION = JHEPA,0305,013;%%
  
  
  
%\cite{Lyth:2005fi}
\bibitem{Lyth:2005fi}
  D.~H.~Lyth and Y.~Rodriguez,
  %``The inflationary prediction for primordial non-gaussianity,''
  Phys.\ Rev.\ Lett.\  {\bf 95} (2005) 121302
  [arXiv:astro-ph/0504045].
  %%CITATION = PRLTA,95,121302;%%
  
  %\cite{GarciaBellido:1995qq}
\bibitem{GarciaBellido:1995qq}
  J.~Garcia-Bellido and D.~Wands,
  %``Metric perturbations in two field inflation,''
  Phys.\ Rev.\ D {\bf 53} (1996) 5437
  [astro-ph/9511029].
  %%CITATION = ASTRO-PH/9511029;%%
  
  
%\cite{Choi:2007su}
\bibitem{Choi:2007su}
  K.~-Y.~Choi, L.~M.~H.~Hall, C.~van de Bruck,
  %``Spectral Running and Non-Gaussianity from Slow-Roll Inflation in Generalised Two-Field Models,''
  JCAP {\bf 0702 } (2007)  029.
  [astro-ph/0701247].


        
\bibitem{VW} F. Vernizzi and D. Wands, JCAP {\bf 0605}, 019
  (2006).[astro-ph/0603799].
  
  %\cite{Byrnes:2009qy}
\bibitem{Byrnes:2009qy}
  C.~T.~Byrnes and G.~Tasinato,
  %``Non-Gaussianity beyond slow roll in multi-field inflation,''
  JCAP {\bf 0908} (2009) 016
  [arXiv:0906.0767 [astro-ph.CO]].
  %%CITATION = ARXIV:0906.0767;%%
  
  %\cite{Battefeld:2009ym}
\bibitem{Battefeld:2009ym}
  D.~Battefeld and T.~Battefeld,
  %``On Non-Gaussianities in Multi-Field Inflation (N fields): Bi and Tri-spectra beyond Slow-Roll,''
  JCAP {\bf 0911} (2009) 010
  [arXiv:0908.4269 [hep-th]].
  %%CITATION = ARXIV:0908.4269;%%
  
 
 %\cite{Alabidi:2006hg}
\bibitem{Alabidi:2006hg}
  L.~Alabidi,
  %``Non-gaussianity for a Two Component Hybrid Model of Inflation,''
  JCAP {\bf 0610} (2006) 015
  [astro-ph/0604611].
  %%CITATION = ASTRO-PH/0604611;%%
  
%\cite{Kim:2010ud}
\bibitem{Kim:2010ud}
  S.~A.~Kim, A.~R.~Liddle and D.~Seery,
  %``Non-gaussianity in axion Nflation models,''
  Phys.\ Rev.\ Lett.\  {\bf 105} (2010) 181302
  [arXiv:1005.4410 [astro-ph.CO]].
  %%CITATION = ARXIV:1005.4410;%%  
  
%\cite{Wang:2010si}
\bibitem{Wang:2010si}
  T.~Wang,
  %``Note on Non-Gaussianities in Two-field Inflation,''
  Phys.\ Rev.\ D {\bf 82} (2010) 123515
  [arXiv:1008.3198 [astro-ph.CO]].
  %%CITATION = ARXIV:1008.3198;%%

%\cite{Peterson:2010mv}
\bibitem{Peterson:2010mv}
  C.~M.~Peterson and M.~Tegmark,
  %``Non-Gaussianity in Two-Field Inflation,''
  Phys.\ Rev.\ D {\bf 84} (2011) 023520
  [arXiv:1011.6675 [astro-ph.CO]].
  %%CITATION = ARXIV:1011.6675;%%

%\cite{Meyers:2010rg}
\bibitem{Meyers:2010rg}
  J.~Meyers and N.~Sivanandam,
  %``Non-Gaussianities in Multifield Inflation: Superhorizon Evolution, Adiabaticity, and the Fate of fnl,''
  Phys.\ Rev.\ D {\bf 83} (2011) 103517
  [arXiv:1011.4934 [astro-ph.CO]];
  %%CITATION = ARXIV:1011.4934;%%
%\cite{Meyers:2011mm}
%\bibitem{Meyers:2011mm}
%  J.~Meyers and N.~Sivanandam,
  %``Adiabaticity and the Fate of Non-Gaussianities: The Trispectrum and Beyond,''
  Phys.\ Rev.\ D {\bf 84} (2011) 063522
  [arXiv:1104.5238 [astro-ph.CO]].
  %%CITATION = ARXIV:1104.5238;%%


%\cite{Elliston:2011dr}
\bibitem{Elliston:2011dr}
  J.~Elliston, D.~J.~Mulryne, D.~Seery and R.~Tavakol,
  %``Evolution of fNL to the adiabatic limit,''
  JCAP {\bf 1111} (2011) 005
  [arXiv:1106.2153 [astro-ph.CO]];
  %%CITATION = ARXIV:1106.2153;%%
%\cite{Elliston:2011et}
%\bibitem{Elliston:2011et}
 % J.~Elliston, D.~Mulryne, D.~Seery and R.~Tavakol,
  %``Evolution of non-Gaussianity in multi-scalar field models,''
  Int.\ J.\ Mod.\ Phys.\ A {\bf 26} (2011) 3821
  [arXiv:1107.2270 [astro-ph.CO]].
  %%CITATION = ARXIV:1107.2270;%%
	
	
%\cite{Byrnes:2010em}
\bibitem{Byrnes:2010em}
  C.~T.~Byrnes and K.~-Y.~Choi,
  %``Review of local non-Gaussianity from multi-field inflation,''
  Adv.\ Astron.\  {\bf 2010} (2010) 724525
  [arXiv:1002.3110 [astro-ph.CO]].
  %%CITATION = ARXIV:1002.3110;%%
	  

	
	
%\cite{Choi:2011me}
\bibitem{Choi:2011me}
  K.~-Y.~Choi and B.~Kyae,
  %``Natural Hybrid Inflation Model with Large Non-Gaussianity,''
  Phys.\ Lett.\ B {\bf 706} (2012) 243
  [arXiv:1109.4245 [astro-ph.CO]].
  %%CITATION = ARXIV:1109.4245;%%
  See also B.~Kyae,
   Eur.\ Phys.\ J.\ C {\bf 72} (2012) 1857    
  %``Twinflation,''  
  [arXiv:0910.4092 [hep-ph]].

	%\cite{Alabidi:2010ba}
\bibitem{Alabidi:2010ba}
  L.~Alabidi, K.~Malik, C.~T.~Byrnes and K.~-Y.~Choi,
  %``How the curvaton scenario, modulated reheating and an inhomogeneous end of inflation are related,''
  JCAP {\bf 1011} (2010) 037
  [arXiv:1002.1700 [astro-ph.CO]].
  %%CITATION = ARXIV:1002.1700;%%
		
	     
\end{thebibliography}
\end{document}